\documentclass[iop]{emulateapj}
\usepackage{amsmath}
\pdfoutput=1

\slugcomment{Not to appear in Nonlearned J., 45.}

\shorttitle{Magnetorotational Instability in Core-Collapse Supernovae}
\shortauthors{Sawai \& Yamada}

\begin{document}

\title{The Evolution and Impacts of Magnetorotational Instability in
  Magnetized Core-Collapse Supernovae} 

\author{Hidetomo Sawai\altaffilmark{1,2} and Shoichi Yamada\altaffilmark{2}} 

\email{hsawai@rist.or.jp}

\altaffiltext{1}{Research Organization for Information Science \&
  Technology, Kobe, Hyogo 650-0047, Japan}
\altaffiltext{2}{Waseda University, Shinjuku, Tokyo 169-8555, Japan}

\begin{abstract}
We carried out 2D-axisymmetric MHD simulations of core-collapse supernovae for
rapidly-rotating magnetized progenitors. By changing both the strength
of the magnetic field and the spatial resolution, the evolution of the
magnetorotational instability (MRI)
and its impacts upon the dynamics are investigated. 
We found that the MRI greatly amplifies the seed magnetic fields in the
regime where not the Alfv\'en mode but the buoyant mode plays a primary
role in the exponential growth phase. 
The MRI indeed has a powerful impact on the supernova dynamics.
It makes the shock expansion faster and the explosion more energetic,
with some models being accompanied by the collimated-jet formations.
These effects, however, are not made by the magnetic
pressure except for the collimated-jet formations.
The angular momentum transfer induced by the MRI causes the expansion
of the heating region, by which the accreting matter gain an additional
time to be heated by neutrinos. The MRI also drifts
low-$Y_p$ matter from the deep 
inside of the core to the heating region, which makes the net neutrino
heating rate larger by the reduction of the cooling due to the
electron capture. 
These two effects enhance the efficiency of the
neutrino heating, which is found to be the key to boost the
explosion. Indeed we found that our models explode far more weakly 
when the net neutrino heating is switched off. The contribution of
the neutrino heating to the explosion energy could reach 60\%
  even in the case of strongest magnetic field in the current simulations. 
\end{abstract}


\section{Introduction}

Magnetic field and rotation are ubiquitous in stars. MiMeS
survey has observed over 550 Galactic O- and B-type stars, and detected
the surface magnetic fields of $\gtrsim 100$~G for $\sim 10$~\% of them (see
\citet{wad14} for review). Estimating the upper limit of the
currently-undetected magnetic
field to be $\sim 100$~G, \citet{wad14} argued
that the distribution 
of the magnetic fields for massive stars may be bimodal: a small
population of strong magnetic fields ($\gtrsim 1$~kG) and a large
majority of weak magnetic fields ($\lesssim 100$~G).
A magnetic field of 1~kG corresponds to the magnetic
flux of $\sim 10^{27}$~G~cm$^{2}$ for a 17~$M_\odot$ star with the
radius of $\sim 8 R_\odot$ \citep{mcn65}, which is comparable to that
of magnetars.  

\citet{ram13}
measured the surface rotational velocities of 216 O-type stars, and
found that 25~\% of the sample have $v \sin i > 200$~km~s$^{-1}$ while
the rest of them are slow rotators. According to stellar evolution
calculations by \citet{woo06}, if a star is rotating fast enough
initially, the rotational mixing prevents a very efficient angular
momentum transport between the helium core and the hydrogen envelope, and
the central iron core maintains a large amount of angular momentum at
pre-collapse stage.
They inferred the rotation period of a neutron
star to be 2.3--9.7~ms for such evolutions of a 16~$M_\odot$ star with
solar metallicity. Then, the high-rotational-velocity 
population found by \citet{ram13} might produce proto-neutron stars
rotating with a 
period similar to those of millisecond pulsars (MSPs).

The influences of magnetic field and rotation on core-collapse
supernovae have been studied as a possible agent to drive explosion
other than neutrino heating, while the latter fails to produce
energetic explosions 
\citep[e.g.,][]{suw10,mue12,bru13}.
MHD core-collapse simulations done so far have placed the main focus on
rather extreme cases, viz., $B_{\textrm{pre}}\sim 10^{12}$--$10^{13}$~G 
and $\Omega_{\textrm{pre}}\sim 1$~rad~s$^{-1}$ at pre-collapse, which
correspond to the magnetar-class magnetic field and MSP-class rotation
\citep[e.g.,][]{yam04,obe06,bur07,shi06,sch08,sch10,saw13a,mos14}.
In those simulations, the 
magnetic field wound by differential rotation grows to dynamically
important strengths and later drives a strong outflow along the
rotation axis, reproducing the typical supernova-explosion energy of
$E_{\textrm{exp}}\sim 10^{51}$~erg. 

Since the magnetic field and rotation in massive stars are
likely to have wide range of values as mentioned above, it may be
also important to study more "ordinary'' cases. In these cases
amplification mechanisms that are more efficient than the 
simple winding are imperative to produce the field strength of $\sim
10^{15}$~G outside the proto-neutron star, which may be necessary to
impact on the supernova dynamics.
For non-rotating case \citet{end10,end12} 
numerically studied the standing accretion shock instability, while
\citet{obe14} investigated the convection. In both cases,
the amplification is rather modest, and the impacts on dynamics
are found to be minor. 

If the iron core is initially rotating rapidly, another candidate of
an efficient field amplification 
mechanism in core-collapse supernovae is the magnetorotational
instability (MRI), which basically occurs in
differentially rotating systems
\citep{bal91,aki03}. Simulations of the MRI for weak seed 
magnetic fields are computationally demanding, since the wavelength of the
fastest growing mode is quite small compared with
the size of the iron core, $\sim 1000$~km:
\begin{eqnarray}
\lambda_{\textrm{FGM}}
&\sim&\frac{2\pi v_{\textrm{A}}}{\Omega}\nonumber\\
&\sim& 200 \textrm{m} 
\left(\frac{\rho}{10^{12}\textrm{g cm}^{-1}}\right)^{-\frac{1}{2}} 
\left(\frac{B}{10^{13}\textrm{G}}\right)
\left(\frac{\Omega}{10^{3}\textrm{rad s}^{-1}}\right)^{-1},\nonumber\\
\end{eqnarray} 
where $v_{\textrm{A}}$ is Alfv\'en velocity\footnote{The wavelength
  of the fastest growing mode given here is one obtained for
  cylindrical rotation laws, $\Omega(\varpi)$, with neglecting
  buoyancy. We deal with the
  general rotation laws, $\Omega(\varpi,z)$, taking the buoyancy into
  account later in Section~\ref{sec.mri}}. 
In fact, most of previous core-collapse simulations assuming
sub-magnetar-class magnetic fields have insufficient spatial
resolutions 
to capture the MRI \citep{moi06,bur07,tak09}\footnote{In spite that
  \citet{moi06} found an exponential growth of magnetic
  field and claimed that the growth is due to a Tayler-type
  "magnetorotational 
    instability'', which is completely different from one found by
    \citet{bal91}. Note however, that the property of the instability
    is still unclear, and no other groups succeeded to reproduce their
    results to date.}.  
In order to resolve the fastest growing mode,
local simulation boxes are utilized in some 2D/3D computations
  \citep{obe09,mas12,gui15b,rem15}. The problems in the local 
simulations, however, are the difficulties in taking into account the
effects from and feedbacks to dynamically changing structures.

\citet{saw13b} conducted the first global core-collapse
simulations for sub-magnetar-class magnetic fields with a sufficient
spacial resolution to capture the MRI albeit in 2D axisymmetry, 
and found that the magnetic field is amplified by the
MRI to dynamically important strengths. In order to study its impacts
on the global dynamical, \citet{saw14} carried out similar but
longer-term simulations up to several hundred milliseconds after
bounce, 
employing the simple light bulb approximation for neutrino
transfer. They found that the MRI 
indirectly enhances the neutrino heating, and thus boost the explosion.
Performing 3D simulations for a thin layer on the equator,
\citet{mas15} argued another possible effect of the MRI, i.e., the
enhancement of neutrino luminosity by MRI-driven turbulence around
the proto-neutron star surface. 

This paper is a sequel to \citet{saw13b} and
\citet{saw14}. We conducted 2D-axisymmetric high-resolution
simulations of core-collapse for rapidly-rotating magnetized
progenitors, changing the initial magnetic field strength and
the spatial resolution. The initial magnetic field strength assumed
here, 
$B_{\textrm{pre}}\sim 10^{11}$~G, are one or two orders of magnitude
smaller than the extreme values adopted in some previous
simulations mentioned above.  

The rest of the paper is organized as follows. In
Section~\ref{sec.method}, we describe the numerical method and
models. The results are presented in Section~\ref{sec.result}, 
and the discussion and conclusion are given in Section~\ref{sec.discon}.

\section{Numerical Methods}\label{sec.method}
We adopt a $15 M_\odot$ star \citep{woo95} for the progenitor of core-collapse simulations,
adding magnetic fields and rotations by hand. The following ideal
MHD equations and the equation of electron number density are
numerically solved by a time-explicit 
Eulerian MHD code, \textit{Yamazakura} \citep{saw13a}:
{\allowdisplaybreaks
\begin{eqnarray}
&&\frac{\partial\rho}{\partial
  t}+\nabla\cdot(\rho\mbox{\boldmath$v$})=0\label{eq.mhd.mass},
\\ 
&&\frac{\partial}{\partial t} (\rho\mbox{\boldmath$v$})+
\nabla\cdot\left(\rho\mbox{\boldmath$v$}\mbox{\boldmath$v$}-
\frac{\mbox{\boldmath$B$}\mbox{\boldmath$B$}}{4\pi}\right)\nonumber\\
&&\hspace{1pc}=-\nabla\left(p+\frac{B^2}{8\pi}\right)-\rho\nabla\Phi 
\label{eq.mhd.mom},
\\  
&&\frac{\partial}{\partial t}\left(e+\frac{\rho
    v^2}{2}+\frac{B^2}{8\pi}\right)
\nonumber\\
&&\hspace{1pc}+\nabla\cdot
\left[
\left(e+p+\frac{\rho
      v^2}{2}+\frac{B^2}{4\pi}\right) 
\mbox{\boldmath$v$}
-\frac{(\mbox{\boldmath$v$}\cdot\mbox{\boldmath$B$}) 
\mbox{\boldmath$B$}}{4\pi}
\right]
\nonumber\\
&&\hspace{4pc}
=-\rho(\nabla\Phi)\cdot\mbox{\boldmath$v$}
+Q_E^{\textrm{abs}}+Q_E^{\textrm{em}}\label{eq.mhd.eng},
\\
&&\frac{\partial\mbox{\boldmath $B$}}{\partial t}=
\nabla\times\left(\mbox{\boldmath$v$}\times\mbox{\boldmath$B$}\right)
\label{eq.mhd.far},
\\
&&\frac{\partial n_e}{\partial
  t}+\nabla\cdot(n_e\mbox{\boldmath$v$})=Q_N^{\textrm{abs}}+Q_N^{\textrm{em}}
\label{eq.ne},
\end{eqnarray}}where $Q_E^{\textrm{abs}}$ and $Q_E^{\textrm{em}}$ are
the changes of energy density due to neutrino/anti-neutrino absorptions 
and emissions, respectively, and  $Q_N^{\textrm{abs}}$ and
$Q_N^{\textrm{em}}$ are the similar notations for the changes of electron
number density. The other symbols have their usual meanings. The electron
fraction, $Y_e$, is given 
by the prescription suggested by \citet{lie05} until bounce. After that,
where Liebend{\"o}rfer's prescription is no longer valid,
Equation~(\ref{eq.ne}) is solved to obtain $Y_e=n_e m_u/\rho$, where
$m_u=1.66\times 10^{-24}$~g is the atomic mass unit. We
assume Newtonian mono-pole gravity. A tabulated nuclear equation of
state produced by \citet{she98a, she98b} is utilized.
Computations are done with polar coordinates in two
dimensions, assuming axisymmetry and equatorial symmetry. 

We take into account interactions of electron
neutrinos $\nu_e$ and anti-neutrinos $\Bar \nu_e$ with nucleons. 
Instead of dealing with detailed neutrino transport, the light bulb
approximation is used as
in \citet{mur09,nor10,han12}. Taking the ultra-relativistic 
limit for electrons and positrons, assuming the Fermi-Dirac distribution
with vanishing chemical potential for $\nu_e$ and $\Bar \nu_e$,
and neglecting the phase space blocking, we evaluate the source
term related to 
$\nu_e$/$\Bar \nu_e$ absorption ($\nu_e+n\to e^-+p, \Bar\nu_e+p\to
e^++n$) in the energy equation (\ref{eq.mhd.eng}) as
\begin{equation}
Q_E^{\textrm{abs}}=
\frac{3\alpha^2+1}{4}
\frac{\sigma_0\langle\epsilon_{\nu_e}^2\rangle}{(m_ec^2)^2}
\frac{\rho}{m_u}
\frac{L_{\nu_e}}{4\pi r^2\langle\mu_\nu\rangle}
\left(Y_n+Y_p\right)\label{eq.qe.abs}
\end{equation} 
\citep{jan01}, where $\alpha=1.26$ is the charged-current axial-vector coupling
constant, $\sigma_0=1.76\times 10^{-44}$~cm$^2$ the characteristic
cross section of weak interaction,
$\langle\epsilon_{\nu_e}^2\rangle=20.8 \left(kT_{\nu_e}\right)^2$ 
the mean square neutrino energy, $m_ec^2=0.511$~MeV the rest-mass
energy of electron, $L_{\nu_e}$ the neutrino luminosity, $r$ the 
distance from the center, and $\langle\mu_\nu\rangle$ the so-called
flux factor. Here, we assume the same luminosity and spectral
temperature for $\nu_e$ and $\Bar \nu_e$. In the present simulations,
$L_{\nu_e}=1.0\times 10^{52}$~erg~s$^{-1}$,
$kT_{\nu_e}=4.0$~MeV, and $\langle\mu_\nu\rangle=1.0$ 
are chosen. Similarly, the source term related to $\nu_e$/$\Bar \nu_e$
absorption in the equation of $n_e$ (\ref{eq.ne}) is 
\begin{equation}
Q_N^{\textrm{abs}}=
\frac{3\alpha^2+1}{4}
\frac{\sigma_0\langle\epsilon_{\nu_e}\rangle}{(m_ec^2)^2}
\frac{\rho}{m_u}
\frac{L_{\nu_e}}{4\pi r^2\langle\mu_\nu\rangle}
\left(Y_n-Y_p\right),\label{eq.qn.abs}
\end{equation}
where $\langle\epsilon_{\nu_e}\rangle=4.11\left(kT_{\nu_e}\right)$
is the mean neutrino energy. 

The source term related to the
$\nu_e$/$\Bar \nu_e$ emission ($e^-+p\to \nu_e+n$, 
$e^++n\to\Bar\nu_e+p$) in Equation~(\ref{eq.mhd.eng}) is given by
\begin{eqnarray}
Q_E^{\textrm{em}}&=&
-\left(3\alpha^2+1\right)
\frac{\pi\sigma_0 c \left(kT\right)^6}{(hc)^3(m_ec^2)^2}
\frac{\rho}{m_u}\nonumber\\
&&\times
\left[Y_n\mathcal{F}_5(-\eta_e)+Y_p\mathcal{F}_5(\eta_e)\right]\label{eq.qe.em}
\end{eqnarray} 
\citep{jan01}, where $\eta_e$ is the electron chemical potential normalized
by the temperature, and
\begin{equation}
\mathcal{F}_l(\eta)\equiv\int^\infty_0 \frac{x^l}{1+\exp(x-\eta)}dx.
\label{eq.fermi}
\end{equation}
Similarly, the emission source term in Equation~(\ref{eq.ne}) is
\begin{eqnarray}
Q_N^{\textrm{em}}&=&
\left(3\alpha^2+1\right)
\frac{\pi\sigma_0 c \left(kT\right)^5}{(hc)^3(m_ec^2)^2}
\frac{\rho}{m_u}\nonumber\\
&&\times
\left[Y_n\mathcal{F}_4(-\eta_e)-Y_p\mathcal{F}_4(\eta_e)\right].\label{eq.qn.em}
\end{eqnarray}
The Fermi integrals (Equation~(\ref{eq.fermi})) are calculated as
follows: 
\begin{eqnarray}
\mathcal{F}_4(\eta_e)&=&24\mathcal{I}_4(\eta_e),\\
\mathcal{F}_5(\eta_e)&=&120\mathcal{I}_5(\eta_e),\\
\mathcal{F}_4(-\eta_e)&=&
 \frac{1}{5}\eta_e^5 + \frac{2}{3}\pi^2\eta_e^3 +
 \frac{7}{15}\pi^4\eta_e + 24\mathcal{I}_4(\eta_e),\\
\mathcal{F}_5(-\eta_e)&=&
 \frac{1}{6}\eta_e^6 + \frac{5}{6}\pi^2\eta_e^4 +
 \frac{7}{6}\pi^4\eta_e^2 
\nonumber\\
&&+ \frac{31}{126}\pi^6 + 120\mathcal{I}_5(\eta_e),
\end{eqnarray}
where 
\begin{equation}
\mathcal{I}_l(\eta)\equiv 
\sum_{m=1}^\infty \left[\frac{(-1)^{m-1}}{m^{l+1}}e^{-m\eta}\right].
\end{equation}

The source terms given by
Equations~(\ref{eq.qe.abs}), (\ref{eq.qn.abs}), (\ref{eq.qe.em}),
(\ref{eq.qn.em}) are valid only in optically thin regions, and must
decrease toward the optically thick regions. To mimic such reduction
they are multiplied by $e^{-\tau_{\textrm{eff}}}$, following \citet{mur09}.
Here, the effective optical depth is defined as
\begin{eqnarray}
\tau_{\textrm{eff}}=\int^\infty_r \kappa_{\textrm{eff}}(r) dr,
\end{eqnarray}
where the effective opacity is given as
\begin{eqnarray}
\kappa_{\textrm{eff}}= 1.2\times10^{-7}
\left(\frac{\rho}{10^{10}\textrm{g cm}^{-3}}\right)
\left(\frac{kT_{\nu_e}}{4\textrm{MeV}}\right)
\left(Y_n+Y_p\right),\nonumber\\
\end{eqnarray}
from the Equations (10), (11), and (14) of \citet{jan01}.

Before conducting high-resolution simulations to capture the
MRI, we first 
follow the collapse of the 15~$M_\odot$ progenitors until several
100~ms after bounce by low-resolution simulations, whose
numerical domain spans from the radius of 100~m to 4000~km. We refer to
these simulations as background (BG) runs. 
In the BG runs, the core is covered with
$N_r\times N_\theta = 720\times 60$ numerical grids, where the spatial
resolution is 0.4--23~km. 

The pre-collapse cores are assumed to be
rapidly rotating with the initial angular velocity profile of 
\begin{eqnarray}
\Omega(r)=\Omega_{0}\frac{r_0^2}{r_0^2+r^2},
\end{eqnarray}
where $r$ is the distance from the center of the core. The parameters
are chosen as $r_0=1000$~km and $\Omega_{0}=2.73$~rad s$^{-1}$,
corresponding to a millisecond proto-neutron star after collapse. 
The initial rotational energy divided by the gravitational binding
energy is $2.5\times 10^{-3}$.

We assume that the pre-collapse magnetic fields have dipole-like
configurations produced by electric currents of a 2D-Gaussian-like 
distribution centered at $(\varpi,z)=(\varpi_0,0)$,
\begin{eqnarray}\label{eq.model.dipole}
j_\phi(\varpi,z)&=&j_0e^{-\tilde{r}^2/2\sigma(\tilde \theta)^2}
\left(\frac{\varpi_0\varpi}{\varpi_0^2+\varpi^2}\right),
\end{eqnarray}
where $(\varpi,z)$ are cylindrical coordinates, $\tilde r \equiv
\sqrt{(\varpi-\varpi_0)^2+z^2}$, $\tilde \theta \equiv 
\arccos (z/\tilde r)$, and 
\begin{equation}
\sigma(\tilde \theta)=
\frac{\tilde r_{\textrm{dec}}}{\sqrt{1-e^2\cos{\tilde \theta}}}
\end{equation}
\citep{saw13a}. Changing $j_0$, we perform
three BG runs with different strengths of magnetic fields, where the
maximum strengths at pre-collapse, $B_{\textrm{pre}}$, are 5.0$\times
10^{10}$, 1.0$\times 
10^{11}$, and $2\times 10^{11}$~G. Hereafter we refer to these
BG runs as B5e10bg, B1e11bg, and B2e11bg, respectively. 
The rest of parameters are set as $\varpi_0=\tilde
r_{\textrm{dec}}=1000$~km and $e=0.5$ in all the computations.
The initial magnetic energy divided by the gravitational binding
energy is quite small, $2.1\times 10^{-6}$, even for the strongest-field
model, B2e11bg. We also computed models without magnetic
field and rotation as well as a model having rotation alone for comparison.

In order to capture the growth of MRI we conduct high-resolution
simulations with the numerical domain spanning $50<(r/$km$)<500$
(referred to as MRI runs). The initial conditions of 
MRI runs are given by mapping the data of the BG runs onto the
above domain at 5~ms after bounce. In order to satisfy the divergence
free constraint on the magnetic field, not the magnetic field itself but
the vector potential is mapped as in \citet{saw13a}. 
The inner and outer radial boundary  
conditions for the MRI runs are given by the data of the basic runs,
except that the inner boundary conditions of $B_r$ are determined to
satisfy the divergence-free condition. The
grid spacing is such that the radial and angular grid sizes are the same, 
viz. $\Delta r=r\Delta\theta$, at the innermost and outermost
cells. For each BG run, four MRI runs with different grid resolutions
are carried out.
Our choice of the resolution
at $r=50$~km, $\Delta_{50}$, (and the numbers of grids, 
$N_r\times N_\theta$), is 12.5~m ($9250\times 6400$), 25~m
($4650\times 3200$), 50~m ($2300\times 1600$), and 100~m ($1160\times
800$). We label the MRI runs by the initial
field strength of the corresponding BG run followed by the
spatial resolution. For 
example, the MRI run using the data of model B5e10bg and
$\Delta_{50}=12.5$~m is referred to as model B5e10$\Delta$12.5. 
For a set of models involving the same initial magnetic
field, we use a term ``model series'', e.g., models series B5e10.

In dealing with the MHD equations in the polar coordinates,
we should be cautious about numerical treatments of
the coordinate singularities at the center of the core ($r=0$) and
the pole ($\theta=0$). In the vicinity of the pole, the regularity
conditions demand that the expansions of $v_{\theta}$, $v_{\phi}$,
$B_{\theta}$, and $B_{\phi}$ with respect to $\theta$ should not contain
$\theta$-independent terms, which is not necessarily
satisfied in numerical simulations. In order to numerically meet the
regularity conditions in the vicinity of the pole albeit
approximately, we remove the region of $\theta<0.3^\circ$ from the
numerical domain and impose boundary conditions based on
the regularity conditions except for $B_{\theta}$, which is determined by
the divergence-free constraint. To diffuse undesirable
fluctuations that tend to violate the regularity, we further introduce
an artificial resistivity only at the cells closest to the pole in the
form of  
\begin{equation}
\eta_{\textrm{a}}= \frac{\alpha a_{\textrm{max}} \Delta^2}{l_B},
\end{equation}
where $\alpha$ is a dimensionless factor, $a_{\textrm{max}}$ the local
maximum characteristic speed, $\Delta$ the grid width, and $l_B$ the
scale height of the magnetic field. 
The factor $\alpha$ is automatically controlled between 0.1--$10^3$
during the simulations depending on how well the regularity condition is
satisfied. 

In order to maintain the regularity conditions approximately around the
center in the BG runs, we remove the central part within the radius of
100~m from the numerical domain and take a similar remedy.

\section{Results}\label{sec.result}
\subsection{The Growth of MRI}\label{sec.mri}
The stability condition of the axisymmetric MRI for general rotation
laws $\Omega(\varpi,z)$ is given by 
\begin{eqnarray}
\mathcal{C}&\equiv&
\left(
  \mathcal{G}_z\mathcal{B}_z\tan^2\theta_k
-2\mathcal{G}_z\mathcal{B}_\varpi\tan\theta_k
+ \mathcal{G}_\varpi\mathcal{B}_\varpi
+ \mathcal{R}_\varpi
\right)
/\Omega^2\nonumber\\
&>&0,
\end{eqnarray}
where $\theta_k$ is the angle between the perturbation wavenumber
$\mbox{\boldmath$k$}$ and the $z$-axis,
\begin{eqnarray}
\mbox{\boldmath$\mathcal{G}$}&\equiv&\frac{\nabla P}{\rho},\\
\mbox{\boldmath$\mathcal{B}$}&\equiv&
-\frac{1}{\Gamma}\frac{\partial\ln P}{\partial s}\Big|_{\rho,Y_e}\nabla s
-\frac{1}{\Gamma}\frac{\partial\ln P}{\partial Y_e}\Big|_{\rho,s}\nabla Y_e,\\
\mbox{\boldmath$\mathcal{R}$}&\equiv&\varpi\nabla\Omega^2,
\end{eqnarray}
and $\Gamma\equiv\partial\ln P/\partial \ln \rho |_{s,Y_e}$ \citep{bal95,obe09}.

The MRI involves two distinct modes, namely, Alfv\'en mode and buoyant
mode, where the former appears for $\mathcal{C}<0$, and the latter
only emerges for $\mathcal{C}+4<0$ \citep{urp96}. Which mode dominates
over the other for a fixed $\theta_k$ depends on the value of
$\mathcal{C}$ \citep{obe09}. For 
$-8<\mathcal{C}<0$, the fastest growing mode is the Alfv\'en mode with
the wavenumber of
\begin{equation}
  \mbox{\boldmath$k$}_{\textrm{FGM}}\cdot \mbox{\boldmath$v$}_{\textrm{A}}=
  \cos\theta_k \Omega\frac{\sqrt{-\mathcal{C}(\mathcal{C}+8)}}{4},
\label{eq.kalf}
\end{equation}
and the growth rate of
\begin{equation}
  \omega_{\textrm{FGM}}=\cos\theta_k \Omega\frac{\sqrt{-\mathcal{C}^2}}{4}.
\end{equation}
For $\mathcal{C}\le-8$, the fastest growth occurs with
\begin{equation}
  \omega_{\textrm{FGM}}=\cos\theta_k \Omega\sqrt{\mathcal{C}+4},
\end{equation}
for 
$\mbox{\boldmath$k$}_{\textrm{FGM}}\cdot\mbox{\boldmath$v$}_{\textrm{A}}=0$,
i.e., it is the buoyant mode. 

\begin{figure}
\epsscale{1}
\plotone{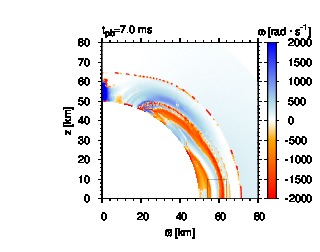}
\plotone{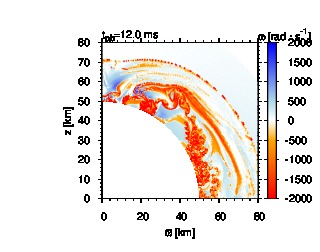}
\caption{Color maps of the dominant modes and growth rate for model
  B5e10$\Delta$12.5 at $t_{\textrm{pb}}=7$ (upper panel) and
  $t_{\textrm{pb}}=12$ (lower panel). The red and blue colors,
  respectively, represent the locations where buoyant mode and Alfv\'en
  mode are dominant. The growth rate is multiplied by -1 for
  buoyant-mode-dominant regions. The boxes in the upper panel
    correspond to the plot areas of Figure~\ref{fig.bpolz}.} 
\label{fig.omri}
\end{figure}

Since $\mathcal{C}$ depends on
$\theta_k$, the dominant mode differs for different directions. In
order to find the dominant mode for a fixed spatial point, 
we vary $\theta_k$ numerically in the range of
$[-\pi/2:\pi/2]$. The result is shown in Figure~\ref{fig.omri} for
model B5e10$\Delta$12.5 at the postbounce time of $t_{\textrm{pb}}=7$
and 12~ms, where the red 
and blue colors represent buoyant-mode- and Alfv\'en-mode-dominant
regions, respectively, and the shades of the colors indicate the
growth rate\footnote{The resolutions of color maps in this paper are
  not the same as those of simulations, where the former are
  reduced to decrease the size
  of figures.}. It is evident that the dominant 
mode is different from 
location to location, and the regions dominated by the buoyant-mode
have on average larger growth rates than those dominated by the
Alfv\'en mode.

\begin{figure}
\epsscale{1.2}
\plotone{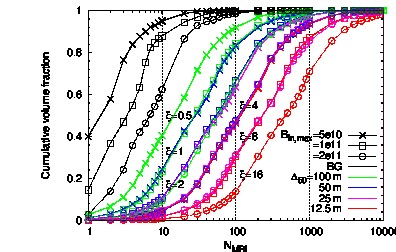}
\caption{Cumulative volume fractions having $N_{\textrm{MRI}}$ smaller
  than a given value for the all models.}
\label{fig.Nmri}
\end{figure}

The growth of the Alfv\'en-mode, although slower than the buoyant mode,
still may have an important effect on 
the magnetic field amplification in the locations of its
dominance. It is hence important to know how 
well we numerically resolve the fastest-growing Alfv\'en mode (FGAM), whose
wave number is given by Equation~(\ref{eq.kalf}). 
Figure~\ref{fig.Nmri} shows for all the models the cumulative fraction
of the volume that has $N_{\textrm{MRI}}$ smaller than a given
value. Here $N_{\textrm{MRI}}$ is defined at each point to be the
ratio of the wavelength of FGAM to the grid size and is measure of how
well FGAM is resolved numerically. 
We introduce to characterize the models, a factor 
$\xi \equiv\left(B_{\textrm{in,max}}/10^{11}\textrm{G}\right)(\Delta_{50}/100
\textrm{m})^{-1}$, since only the initial
strength of the magnetic field and spacial resolution are different
among the current set of the models.
In fact, similar distributions is obtained for the
models having the same $\xi$ at early epochs when the magnetic field
is almost passive (see Figure~\ref{fig.Nmri}). According to  
\citet{shi06}, $N_{\textrm{MRI}}\gtrsim 10$ is required to capture the
linear growth of the Alfv\'en mode. In our weakest-field model series
B5e10, the volume fractions with $N_{\textrm{MRI}}\le 10$ are 0.6,
0.27, 0.12, and 0.018 for models $\Delta$100 ($\xi=0.5$), $\Delta$50
($\xi=1$), $\Delta$25 ($\xi=2$), and $\Delta$12.5 ($\xi=4$),
respectively.
We hence believe that our highest-resolution models should be able to
capture the linear growth of the Alfv\'en mode well.

\begin{figure*}
\epsscale{1}
\plottwo{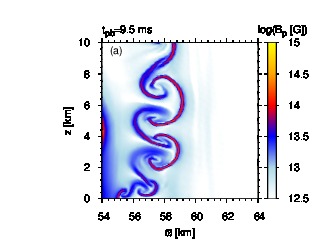}{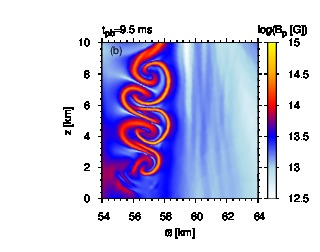}
\plottwo{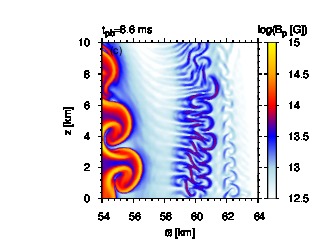}{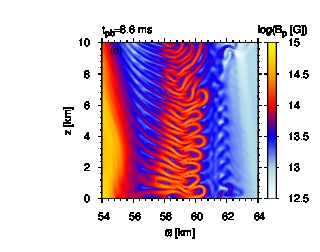}
\plottwo{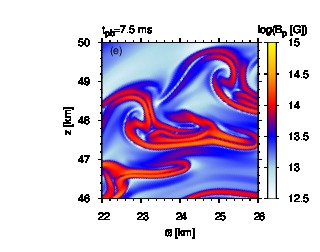}{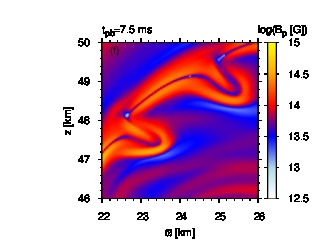}
\caption{Color maps for the strength of the poloidal magnetic field
  for models B5e10$\Delta$12.5 (left panels) and B2e11$\Delta$12.5
  (right panels). The upper four panels zoom in a part of equatorial
  region (presented by the large box in the upper panel of
  Figure~\ref{fig.omri}), while the lower two panels that of a
  middle-latitude  region (the small box in the upper panel of
  Figure~\ref{fig.omri}). Panels (c) and (d) are for models with
  initial perturbations.} 
\label{fig.bpolz}
\end{figure*}

\begin{figure}
\epsscale{1}
\plotone{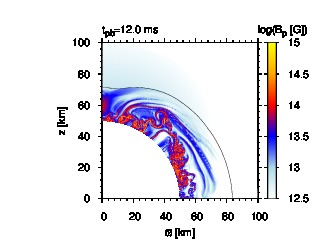}
\plotone{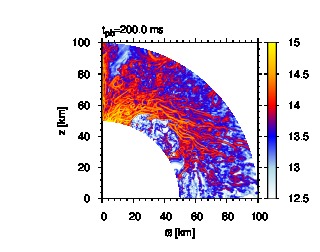}
\caption{Color maps for the strength of the poloidal magnetic fields
  for model B5e10$\Delta$12.5 at
  $t_{\textrm{pb}}=12$~ms (upper panel) and 200~ms (lower panel). The
  black line in the upper panel represents the shock surface.} 
\label{fig.bpol}
\end{figure}

In order to confirm that both the buoyant and Alfv\'en modes are
growing indeed in the regions predicted in Figure~\ref{fig.omri}, we
examine the wavelengths of the growing modes in these regions.
The upper panels of Figure~\ref{fig.bpolz}
show the color maps of the poloidal magnetic field
strength at $t_{\textrm{pb}}=9.5$~ms in a region around the
equator, indicated by the large box in the upper panel of 
Figure~\ref{fig.omri}, where two red belts of buoyant-mode dominance are
observed\footnote{Although the upper panel of 
  Figure~\ref{fig.omri} is depicted for model B5e10$\Delta$12.5, its
  feature is very similar among all the models at this point
  of time.}. 
 We found that the patterns of strong-magnetic-field filaments
   seen in the upper panels of Figure~\ref{fig.bpolz} have grown in
   the regions of buoyant-mode dominance: in the case of model
   B5e10$\Delta$12.5, this region corresponds to the right-side red
   belt observed in the large box indicated in the upper panel of
   Figure~\ref{fig.bpol}, which has been advected leftward  
   during $t_{\textrm{pb}}=7.0$--$9.5$~ms.
We compare models B5e10$\Delta$12.5 and B2e11$\Delta$12.5, which have
different initial field strengths.
As expected for the buoyant mode, the wavelengths of the growing
modes, which are evaluated from the sizes of the patterns, are nearly
identical between the two models.
The wavelengths of
the growing modes observed here may reflect the scale of the dominant
perturbation, which may come from numerical noises. 
To see if this is true, we performed test simulations for the two
models in which  
a perturbation of $u'=u'_0\sin (2\pi z/\lambda_{\textrm{prt}})$ is
given at the beginning of the MRI runs, where the amplitude and
the wavelength of the perturbation is set as 1\% and 500~m,
respectively. As a result, we found that the wavelengths of
growing modes are 
shorter than those of the models without perturbation
(see panels (c) and (d) of 
Figure~\ref{fig.bpolz}), which indicates that the observed modes
depend on the dominant scale of perturbations. 

The lower panels of Figure~\ref{fig.bpolz}
zoom in the area around $\theta=35^\circ$ in the vicinity of the
inner boundary, where a pocket of buoyant-mode-dominant regions are
surrounded 
by an Alfv\'en-mode-dominant region (see the small box in the
upper panel of 
Figure~\ref{fig.omri}). The wavelength of the growing mode there is
shorter for the weaker initial field. In fact, the
widths of protruding magnetic flux loops in the panel (e) of
Figure~\ref{fig.bpolz} for model B5e10$\Delta$12.5 
are about three times smaller than those in the panel (f) for model
B2e11$\Delta$12.5. The ratio is close to four, the value expected for
the Alfv\'en mode. With these facts, we believe that our simulations
capture both the buoyant and Alfv\'en modes correctly. 

\begin{figure}[h]
\epsscale{1}
\plotone{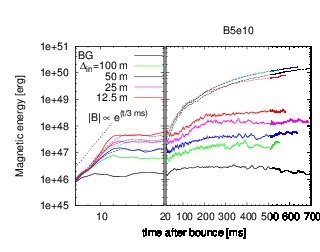}
\plotone{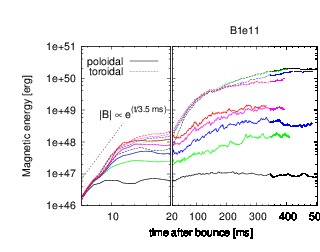}
\plotone{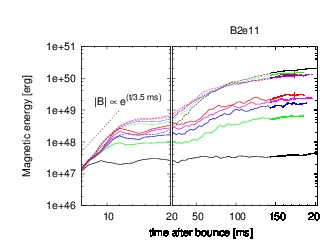}
\caption{Evolutions of the magnetic energies of the poloidal
  and toroidal components, integrated over the whole numerical domain,
  for model series B5e10 (upper panel), B1e11 
  (middle panel), and B2e11 (lower panel). Those for the MRI
    runs are plotted until the shock surface reaches the outer
    boundary in each model. Only those for model B2e11$\Delta$12.5 are
    plotted even after the shock passes the outer boundary in order to
    see a clear saturation of the magnetic energy of the poloidal
    component. The red crosses represent the moment when the shock
    surface reaches the outer boundary in model B2e11$\Delta$12.5.}
\label{fig.t-emag}
\end{figure}

Figure~\ref{fig.t-emag} plots the time evolutions of the magnetic
energies, integrated over the whole numerical domain for all
models. Those for the MRI runs are plotted until the shock surface
reaches the outer boundary. Only in model B2e11$\Delta$12.5, the plots 
are continued for another 12~ms after the shock surface passes the
outer boundary. Since the magnetic energies flowing out of the
boundary during this 12~ms are found to be negligible, i.e., 0.81\%
and 0.24\% of the total poloidal and toroidal magnetic energies at the
end of the plot, respectively, we do not take them into account in the
following discussion.
The exponential growth of the energy of the 
poloidal component, $E_{B_{\textrm{p}}}$, is apparent during the first
$\sim 10$~ms for all the MRI runs. In 
each model series, the growth timescale becomes shorter (or the growth rate
is larger) for higher resolutions until it converges to
$\approx$3--3.5~ms. These timescales well match the theoretical
prediction for the buoyant-mode of $\sim$2000~rad~s$^{-1}$, which is
shown in the upper panel of  
Figure~\ref{fig.omri}. This implies that the exponential growth is
dominated by the buoyant mode. Indeed, the comparison between the
lower panel of Figure~\ref{fig.omri} and the upper panel of
Figure~\ref{fig.bpol} indicates the coincidence of the locations,
where the poloidal magnetic field is preferentially amplified, with
those of buoyant-mode dominance at 
$t_{\textrm{pb}}=12$~ms, around the end of the exponential growth.
From the numerical convergence we observed, it is suggested that the
high spatial resolution is required even for the buoyant mode,
in which all wavelengths grow at an equal rate.

After the exponential growth phase ceases, $E_{B_{\textrm{p}}}$
continues to increase gradually until it reaches 
saturation roughly around $t_{\textrm{pb}}=$210, 270, and 160~ms for
model series B5e10, B1e11, and 
B2e11, respectively (see Figure~\ref{fig.t-emag}). During this phase
the region of strong magnetic field, say, $B>10^{14}$~G,
spreads over a considerable volume inside the radius of $\sim
100$~km (see the lower panel of Figure~\ref{fig.bpol} for B5e10$\Delta$12.5). 

\begin{figure}
\epsscale{1}
\plotone{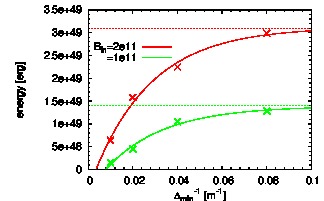}
\caption{The time-averaged saturation values of the magnetic energy of
  the poloidal component (crosses) and the fitted curves (solid
  lines) with respect 
to the resolution for model series B1e11 and B2e11. The dotted lines
represent the saturation values, $a_1$.}
\label{fig.bsat}
\end{figure}

As can be seen from each panel of Figure~\ref{fig.t-emag}, the
saturated values of $E_{B_{\textrm{p}}}$ do
not converge, which may be because the turbulence is not yet fully
captured due to numerical diffusivity
\citep{saw13b}. Nevertheless, since model
series B1e11 and B2e11 show a trend of convergence, we may be
able to estimate the converged values by fitting the time-averaged
$E_{B_{\textrm{p}}}$ for the different resolution models with suitable
functions.  Taking the time
averages over $t_{\textrm{pb}}=270$--330~ms 
and 165--185~ms for model series B1e11 and B2e11,
respectively, we fitted the results with functions in the form of 
\begin{equation}
\langle E_{B_{\textrm{p}},{\textrm{sat}}}\rangle
=a_1-a_2\exp\left({-a_3/\Delta_{50}}\right),
\end{equation}
where $a_1$, $a_2$, and $a_3$ are the parameters to be determined.
The values obtained for $a_1$ are
$1.4\times 10^{49}$ and $3.1\times 10^{49}$~erg for model series B1e11
and B2e11, respectively (see Figure~\ref{fig.bsat}). It is also found
that the saturated values of $E_{B_{\textrm{p}}}$ for
the highest-resolution models B1e11$\Delta$12.5 and B2e11$\Delta$12.5
are 91\% and 96\% of these values, respectively, i.e., they
are close to  
convergence. Indeed, it is expected that if we were
able to afford twice higher resolution, we could achieve convergence.

Our results also suggest that larger initial
magnetic fields may result in larger saturated values. This is
consistent with the results obtained by \citet{haw95}, who performed
local box simulations of MRI in the context of accretion
disks. \citet{mas15} also claimed that the saturation depends on the
initial fields, however, no resolution study was done. As shown here, the 
resolution dependence should properly taken into account in discussing
the saturation. 

The magnetic energy of the toroidal component, $E_{B_{\textrm{t}}}$,
also shows the exponential growth in each model (see
Figure~\ref{fig.t-emag}). Since the non-axisymmetric MRI for the
toroidal components cannot be treated with the current 
simulations, this is not due to the MRI but due to the winding 
of the MRI-amplified poloidal component by differential rotation.
This is understood from the fact that $E_{B_{\textrm{t}}}$
continuously increases even after the poloidal component is saturated
and become one order of magnitude greater 
than $E_{B_{\textrm{p}}}$. At the end of the simulations, 
$E_{B_{\textrm{t}}}$ still continues to increase gradually in most of
the models. Only for models B1e11$\Delta$25 and B1e11$\Delta$12.5,
$E_{B_{\textrm{t}}}$ has nearly reached the saturated values.
Incidentally, the numerical convergence is achieved in
$E_{B_{\textrm{t}}}$ for model series B1e11 and B2e11.

\subsection{Impacts on Global Dynamics}\label{sec.dyn}
\subsubsection{Background Runs}
\begin{figure}
\epsscale{1}
\plotone{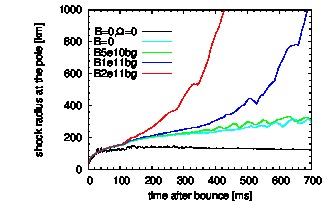}
\caption{Time evolutions of the maximum shock radii for BG runs.}
\label{fig.t-rsh.bg}
\end{figure}

With our choice of $L_\nu=1.0$~erg~s$^{-1}$, the BG run with
no magnetic field and rotation fails to explode,
the shock wave being stalled at $r\lesssim 150$~km (see black line in
Figure~\ref{fig.t-rsh.bg}). 
Although the shock surface is deformed by SASI-like 
oscillations during the early postbounce phase, which is imprinted in
the zigzag evolution of the shock radius in
Figure~\ref{fig.t-rsh.bg} at $t_{\textrm{pb}}\lesssim 300$~ms,
it becomes almost spherically symmetric later on.

Initial rotation of $|T/W|=0.25$~\%
substantially changes the behavior of the shock
evolution. In fact, fluids at middle to low latitudes tend to
expand toward a larger 
radius thanks to centrifugal forces. The
maximum shock radius gradually 
increases and exceeds 200~km by $t_{\rm{pb}}=700$~ms (see the cyan line in
Figure~\ref{fig.t-rsh.bg}), at which time some parts of fluid elements
are still going outward albeit slowly. These features are
consistent with the former findings \citep{suw10,nak14,iwa14},
that the rotation helps the explosion (See, however, \citet{mar09}). 

In the BG runs with both magnetic field and rotation, the shock
surface propagates 
outward more easily compared with the rotation-only model, with
faster propagation speeds for stronger initial magnetic fields
(Figure~\ref{fig.t-rsh.bg}). The panels (a), (b), and (c) of
Figure~\ref{fig.betamulti} show that the regions of low plasma beta
($\beta$, the ratio of matter pressure to magnetic pressure) 
appear around the mid-latitude, indicating that
the magnetic pressure plays an important role to push the shock outward.

\subsubsection{Dynamical Behavior of MRI Runs}

\begin{figure*}
  \begin{minipage}{0.33\hsize}
    \includegraphics[scale=0.55]{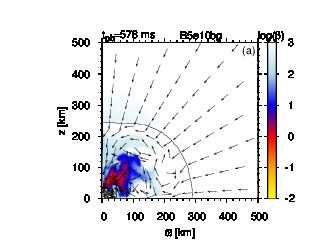}
  \end{minipage}
  \begin{minipage}{0.33\hsize}
    \includegraphics[scale=0.55]{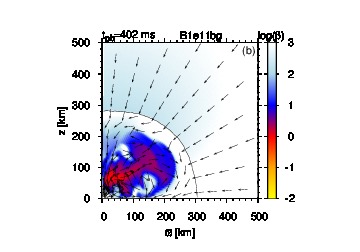}
  \end{minipage}
  \begin{minipage}{0.33\hsize}
    \includegraphics[scale=0.55]{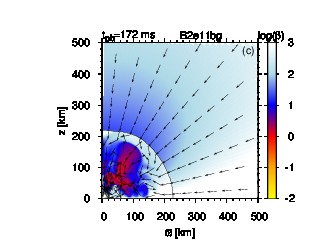}
  \end{minipage}

  \begin{minipage}{0.33\hsize}
    \includegraphics[scale=0.55]{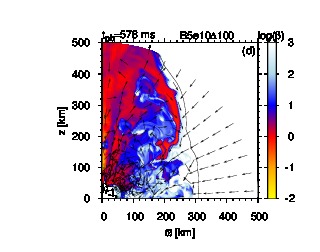}
  \end{minipage}
  \begin{minipage}{0.33\hsize}
    \includegraphics[scale=0.55]{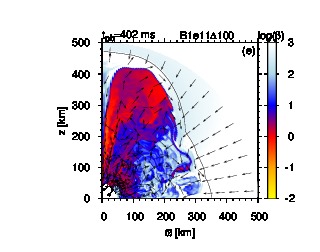}
  \end{minipage}
  \begin{minipage}{0.33\hsize}
    \includegraphics[scale=0.55]{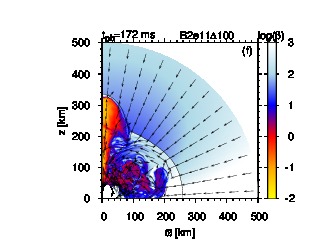}
  \end{minipage}

  \begin{minipage}{0.33\hsize}
    \includegraphics[scale=0.55]{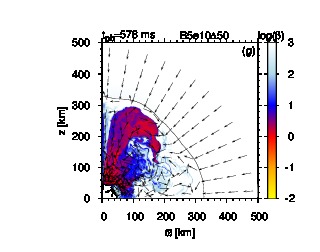}
  \end{minipage}
  \begin{minipage}{0.33\hsize}
    \includegraphics[scale=0.55]{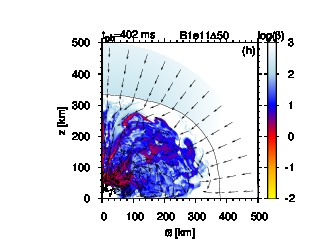}
  \end{minipage}
  \begin{minipage}{0.33\hsize}
    \includegraphics[scale=0.55]{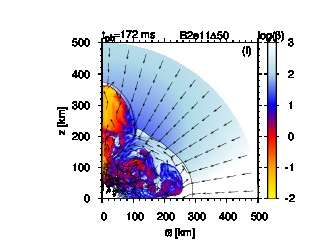}
  \end{minipage}

  \begin{minipage}{0.33\hsize}
    \includegraphics[scale=0.55]{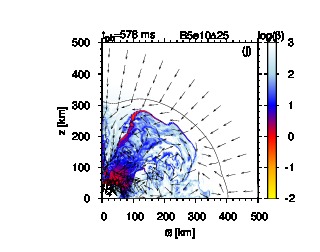}
  \end{minipage}
  \begin{minipage}{0.33\hsize}
    \includegraphics[scale=0.55]{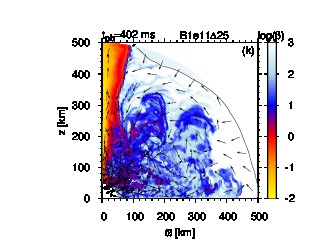}
  \end{minipage}
  \begin{minipage}{0.33\hsize}
    \includegraphics[scale=0.55]{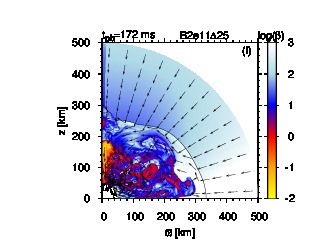}
  \end{minipage}

  \begin{minipage}{0.33\hsize}
    \includegraphics[scale=0.55]{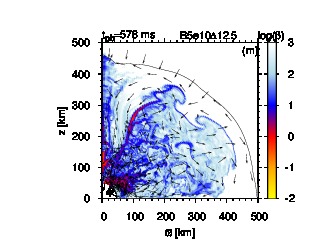}
  \end{minipage}
  \begin{minipage}{0.33\hsize}
    \includegraphics[scale=0.55]{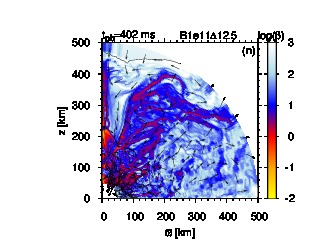}
  \end{minipage}
  \begin{minipage}{0.33\hsize}
    \includegraphics[scale=0.55]{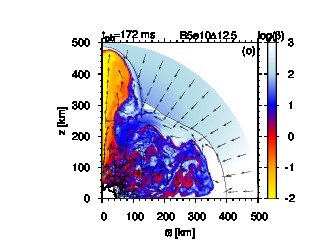}
  \end{minipage}
  \caption{Color maps of the plasma beta with velocity vectors
    presented by arrows at $t_{\textrm{pb}}=$578, 402,
  and 172~ms for model series B5e10, B1e11, B2e11, respectively.}
  \label{fig.betamulti}
\end{figure*}

\begin{figure*}[t]
\plottwo{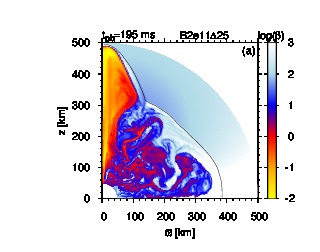}{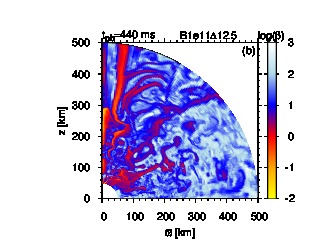}
  \caption{Color maps of the plasma beta for model
      B2e11$\Delta$25 at $t_{\textrm{pb}}=$195~ms (a) and model
      B1e11$\Delta$12.5 at $t_{\textrm{pb}}=$440~ms (b), which are
      supplemental plots for panels (l) and (n) of
      Figure~\ref{fig.betamulti}, respectively.}
\label{fig.beta2}
\end{figure*}

The dynamics change even more drastically when the spatial
resolution is increased (MRI runs).
Figure~\ref{fig.betamulti} displays the distributions of the plasma
beta for all the 15 models at $t_{\textrm{pb}}= 578$, 402,
and 172~ms for model 
series B5e10 (left column), B1e11 (middle column), and B2e11
(right column), respectively. In model series B2e11, all the MRI runs
result in the formation of a collimated low-$\beta$ jet emerging
from inside the roughly-spherical shock, whereas the BG run yields an
almost spherical expansion of the shock wave. Note that the
  low-$\beta$ region seen in panel (l) for model B2e11$\Delta$25 at
  172~ms evolves into a collimated jet later on as shown in panel
  (a) of Figure~\ref{fig.beta2}. Meanwhile, in 
a weaker-field model series B1e11, the situation is
not as simple as in model series B2e11. As the resolution gets higher
from model B1e11bg to 
B1e11$\Delta$100, the shape of shock surface changes from spherical to
prolate, but it returns to spherical shape when the resolution is
doubled again
(see panels (b), (e), (h) of Figure~\ref{fig.betamulti}). Another
doubling of the resolution, in 
turn, brings about the formation of a collimated low-$\beta$
jet emerging from inside the spherical shock (model
B1e11$\Delta$25, panel (k)). In the highest resolution 
model B1e11$\Delta$12.5, a low-$\beta$ region is observed
  around the radius of 200~km in the vicinity of the pole (panel (n) of
  Figure~\ref{fig.betamulti}), which may hint at a later jet
  formation. Although the head of low-$\beta$ region is still lingering
  around the radius of 300~km at the end of the simulation
  ($t_{\textrm{pb}}=$440~ms, see panel (b) of Figure~\ref{fig.beta2}),
  we expect that it would propagate further and eventually forms a
  collimated jet as found in model B1e11$\Delta$25 (see below).
Finally in the weakest-field model series B5e10, the
shock surfaces are roughly spherical for all the resolutions except
model B5e10$\Delta$100, which has a prolate shock, and no model shows
a jet formation until the end of the simulation. 

\begin{figure*}[t]
\epsscale{1}
\plottwo{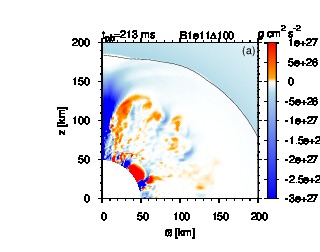}{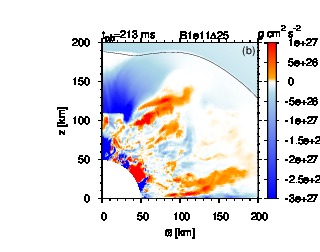}
\plottwo{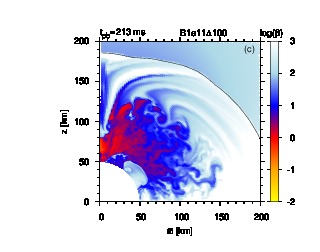}{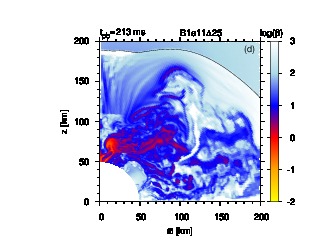}

\caption{Color maps of the ram pressure (upper panels) and the plasma
beta (lower panels) for model B1e11$\Delta$100 (left panels) and
B1e11$\Delta$25 (right panels). The ram pressure is multiplied by
$-1$ where the radial velocity is negative.}
\label{fig.rambeta}
\end{figure*}

We first discuss the factors responsible for the different shock
morphorogies for different resolutions by comparing
models B1e11$\Delta$100 and B1e11$\Delta$25 at
$t_{\textrm{pb}}=213$~ms, several milliseconds 
prior to the launch of the collimated-jet in model B1e11$\Delta$25.
The upper panels of Figure~\ref{fig.rambeta} depict the distribution
of the ram pressure for the two models. It is observed in both the 
models that a vicinity of the pole is dominated by intense downflow
(blue region), outside of which modest outflow driven by
relatively-low plasma beta is seen (red region; see also lower panels). 
The width of the downflow channel is found to be narrower for the lower
resolution model B1e11$\Delta$100, which would be due to less
effective MRI: the better the MRI resolved, the more efficiently
angular momentum is transferred outwards from the rotation axis, with
which the rotational support decreases further around the pole and
a broader downflow channel forms. Accordingly, the lower-$\theta$ 
edge of the outflow gets closer to the pole in this
model, viz. the matter is ejected more preferentially along the pole.
It is likely that this causes the prolate shock surface found in a model
B1e11$\Delta$100. Note that although relatively-low plasma beta,
$\beta\sim 1$, is seen around the bottom of the downflow channel in
both the models, it seems not enough to drive the matter outward
against the downflow (see the lower panels of
Figure~\ref{fig.rambeta}). The magnetically-driven mass ejection is
only possible for the region outside the channel with such
relatively-low plasma beta. It is found from Figure~\ref{fig.betamulti}
that the trend of broader downflow channel and thus a larger
deflection of the outflow direction from the 
pole for higher resolution models is valid for a wide range of
resolution in model-series B5e10 and B1e11 as long as a jet is
absent. This suggest that our interpretation for the shock morphology
is reasonable. Note that the trend discussed here becomes no longer
valid once a jet appears, since it changes the flow structure.

\begin{figure*}[t]
\epsscale{1}
\plottwo{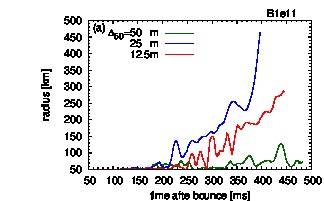}{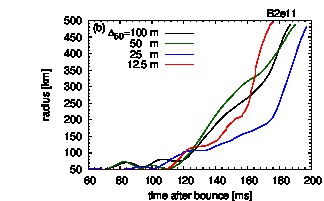}
\plottwo{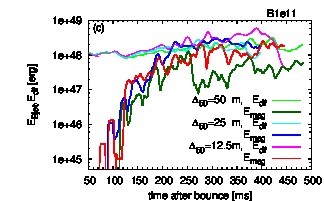}{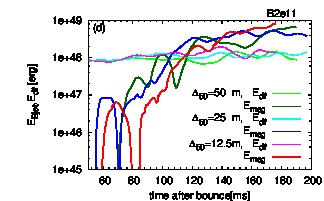}
\caption{Evolution of low-$\beta$ head (upper panels), and the
    downflow energy, $E_{\textrm{df}}$ and magnetic energy
    responsible for jet driving, $E_{B_\textrm{jet}}$ (lower
    panels). The left and right columns are for models series B1e11
    and B2e11, respectively. Note that the time average over the
    interval of 10~ms is taken for each plot.}
\label{fig.t-jet}
\end{figure*}

According to the above discussion, the collimated-jets seen in some
models must have been launched against the downflow. 
In the case of model B1e11$\Delta$25, the low-$\beta$ clump around
$r=70$~km in the vicinity of the pole observed in panel (d) of
Figure~\ref{fig.rambeta} is a prototype of the low-$\beta$ jet seen
at a later phase (panel (k) of Figure~\ref{fig.betamulti}). We found
indeed that this low-$\beta$ clump suffers from successive depression by the
downflow until it finally forms the collimated jet.

To see the process of jet formation more in detail, we define the
  ``low-$\beta$ head'' by the maximum radius of the region where the
  ratio of the matter 
  pressure to magnetic pressure, each of which is angularly averaged
  over $\theta\le5^\circ$\footnote{The average of variable A is
    taken as $\int A dV/\int dV$.}, is less than 0.5, and plot the
  evolutions of 
  them for model series B1e11 and B2e11 (see the upper panels of
  Figure~\ref{fig.t-jet}). For model B1e11$\Delta$25 the sequence of
  the depression described above is clearly seen as an oscillatory
  evolution of the low-$\beta$ head until $t_{\textrm{pb}}\approx
  370$~ms, from which it grows monotonically. Model
  B1e11$\Delta$12.5 shows a slower growth of 
  the low-$\beta$ head and more distinct feature of oscillation,
  indicating that the depression by downflow is more significant. A
  similar oscillation is also found for model B1e11$\Delta$100, but
  the low-$\beta$ head almost stagnates at a small radius in this
  case. On the contrary to the weaker field case, model series B2e11
  shows no oscillation of the low-$\beta$ head, which grows almost
  monotonically in all of the three models plotted in panel (b) of
  Figure~\ref{fig.t-jet}\footnote{As shown, the low-$\beta$ heads
      evolve more or less similarly among all the MRI runs
      of model series B2e11. Although, in the right column of
      Figure~\ref{fig.betamulti}, all the MRI runs except for
      B2e11$\Delta$25 show a trend that the jet-head 
      radius is larger for a higher resolution at 172~ms, with which one may
      think that model B2e11$\Delta$25 is an outlier,
      Figure~\ref{fig.t-jet} represents that there is
      in fact no such a trend on the whole time.}.

From the above discussion, the downflow seems the
  key to the propagation of low-$\beta$ head and the eventual formation of a
  collimated jet. Then the condition for the jet formation may be
  obtained by comparing the kinetic energy of the downflow and the magnetic
  energy responsible for the jet driving. As argued below, we indeed
  found that this comparison reasonably explains the jet formation.

In the lower panels of
  Figure~\ref{fig.t-jet} we plot the time evolutions of the above two
  energies for model series B1e11 and B2e11, defining the former as
$E_{\textrm{df}}\equiv\int_{v_r<0}(\rho v_r^2/2)dV$ and 
the latter as $E_{B_\textrm{jet}}\equiv\int_{\beta<0.5}(B^2/8\pi) dV$,
where the integrants are 
nonzero only for $v_r<0$ and $\beta<0.5$, respectively, and the
integration ranges are confined to 50~km$\le r \le r_{\textrm{sh}}$ and
$\theta<20^\circ$. In model series B2e11, the evolutions of
$E_{B_\textrm{jet}}$ appears rather similar among the different
resolutions. In each model, $E_{B_\textrm{jet}}$ exceeds
$E_{\textrm{df}}$ around $t_{\textrm{pb}}\approx 100$~ms, which is found
to approximately coincide with the start of the low-$\beta$ head
propagation (see right column of Figure~\ref{fig.t-jet}). This
suggests that the $E_{\textrm{df}}$-$E_{B_\textrm{jet}}$ comparison is
indeed a rough indicator for the jet formation.

\begin{figure}
\epsscale{1}
\plotone{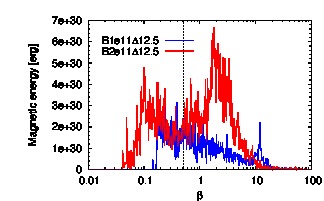}
\caption{$\beta$-distribution of magnetic energy contained within
  $\theta<20^\circ$ for models B1e11$\Delta$12.5 and B2e11$\Delta$12.5
  at the moment when $E_{\textrm{df}}$ first reaches $2.0\times
  10^{48}$~erg in each model, $t_{\textrm{pb}}=198$~ms for the former and
  $t_{\textrm{pb}}=140$~ms for the latter. The vertical-dotted line
  represents $\beta=0.5$.}
\label{fig.beta-emag}
\end{figure}

Unlike in model series B2e11, the values of
  $E_{B_\textrm{jet}}$ in model series B1e11 rather diverge among
  different resolutions during a late phase after
  their growth nearly saturates around $t_{\textrm{pb}}\approx
  250$~ms. Interestingly, while reducing the grid size, $\Delta_{50}$,
  from 50~m to 25~m result in averagely-larger $E_{B_\textrm{jet}}$
  during the late phase, which may be simply due to smaller numerical
  diffusivity, another doubling of resolution decreases that value
  (see panel (c) of Figure~\ref{fig.t-jet}). We infer that the
  decrease of $E_{B_\textrm{jet}}$ observed here is caused by the
  interaction of low-$\beta$ matter with downflow. Since the downflow
  matter involves high-$\beta$ (see Figure~\ref{fig.rambeta}), the
  plasma beta of low-$\beta$ matter increases as it hit by and mixed
  with the downflow matter, which results in downturn of
  $E_{B_\textrm{jet}}$. Indeed, the panel (c) of
  Figure~\ref{fig.t-jet} shows that the downflow energy,
  $E_{\textrm{df}}$, is averagely larger in model B1e11$\Delta$25 than
  in B1e11$\Delta$12.5, which is due to more effective MRI as
  discussed before, and the effect of downflow is expected to be more
  standout for the latter model. Although the effect of the downflow
  basically becomes more potent by increasing the resolution, whether
  it is essential for the change of $E_{B_\textrm{jet}}$ would depend
  on the competition with other factors. For the increase of
  $E_{B_\textrm{jet}}$ from model B1e11$\Delta$50 to B1e11$\Delta$25,
  it is likely that the reduction of the numerical diffusivity is more
  important than the increment of the downflow effect. Meanwhile, the
  fact that the $E_{B_\textrm{jet}}$ is roughly unchanged by
  increasing the resolution in model series B2e11 indicates that the
  downflow effect is insignificant in these models. One reason
    for this would be that
  the plasma beta of low-$\beta$ matter is low enough to maintain
  $\beta < 0.5$, the criterion for adding up $E_{B_\textrm{jet}}$, even
  after the mixing with downflow matter. We compare in
    Figure~\ref{fig.beta-emag} the
    $\beta$-distribution of magnetic energy contained within $\theta<20^\circ$
    for models B1e11$\Delta$12.5 and B2e11$\Delta$12.5 at the moment
    when $E_{\textrm{df}}$ first reaches $2.0\times 10^{48}$~erg in
    each model, and
    found that the latter model indeed involves more low-$\beta$
    matter. Another reason that we consider important is that models
    B2e11 take shorter 
    time to reach the saturation of magnetic energy than models B1e11
    do (see Figure~\ref{fig.t-emag} and lower panels of
    Figure~\ref{fig.t-jet}). Since $E_{\textrm{df}}$ gradually
    increases until attenuated by the jet formation, an early
    growth of magnetic energy is advantageous to alleviate the downflow
    effect. In fact, the value of $E_{\textrm{df}}$ at the
    moment when it is caught up with by $E_{B_\textrm{jet}}$, is
    generally smaller in model 
    series B2e11 ($E_{\textrm{df}}=0.6-1.2\times 10^{48}$~erg) than in B1e11
    ($E_{\textrm{df}}=1.5-2.3\times 10^{48}$~erg).

Bearing in mind the variations of $E_{B_\textrm{jet}}$ and
  $E_{\textrm{df}}$ among the different resolutions mentioned
  for model series B1e11 in the above, the non-monotonic dependence of jet
  formation on the resolution found in these models (the middle column
  of Figure~\ref{fig.betamulti}) is also explained reasonably in terms
  of the $E_{\textrm{df}}$-$E_{B_\textrm{jet}}$ comparison. For model
  B1e11$\Delta$50, the fact that $E_{B_\textrm{jet}}$ is almost always
  smaller than $E_{\textrm{df}}$ is consistent with the stagnation of
  the low-$\beta$ head at small radii. Similar to 
  model series B2e11, model B1e11$\Delta$25 shows the outward
  propagation of the low-$\beta$ head after $E_{B_\textrm{jet}}$
  becomes comparable to $E_{\textrm{df}}$ around
  $t_{\textrm{pb}}\approx 220$~ms. Contrary to the former cases,
  however, $E_{B_\textrm{jet}}$ does not exceed $E_{\textrm{df}}$ so
  much and sometimes even falls behind that, as expected from the
  oscillatory evolution of the low-$\beta$ head. In the higher
  resolution model B1e11$\Delta$12.5, the low-$\beta$ head starts to
  propagate after $E_{B_\textrm{jet}}$ grows comparable to
  $E_{\textrm{df}}$ around $t_{\textrm{pb}}\approx 280$~ms, but shows
  remarkable oscillations as $E_{B_\textrm{jet}}$ occasionally becomes
  smaller than $E_{\textrm{df}}$ by up to factor $\approx 10$, due to
  the downflow effect. About 140~ms later, 
  however, low-$\beta$ filaments outside the downflow choke the
  channel region (see panel (b) of Figure~\ref{fig.beta2}), decreasing
  $E_{\textrm{df}}$ drastically and resulting in the acceleration of
  the low-$\beta$ head. As mentioned before the position of the
  low-$\beta$ head is still at $r<300$~km and no clear jet formation
  is observed by the end of the simulation. Nevertheless, since
  $E_{B_\textrm{jet}}$ exceeds $E_{\textrm{df}}$ by almost factor 10
  at that time, and the latter does not increase significantly 
  afterward, we expect that a collimated jet will form later also in
  model B1e11$\Delta$12.5.
  It should be noted that since doubling the resolution from model
  B1e11$\Delta$25 to B1e11$\Delta$12.5 renders the downflow effect
  more significant,
  which is disadvantageous for a jet formation, higher resolution runs
  are necessary to understand how the dynamics converge in 
  terms of resolution for model series B1e11.

Since the jet formations discussed above take place close to
  the pole, where the coordinates become singular, one may be worried
  that the observed features are merely numerical artifacts. Although
  some level of numerical noises originating from the coordinate
  singularity may be inevitable in spite of the special treatment
  described in Section~\ref{sec.method}, we believe that they are of physical
  origin. This is because the jet is born at some distance from the
  pole, $\approx 10$~km (panel (d) of Figure~\ref{fig.rambeta}), which
  is much larger than the width of 
  the region of the special treatment, and because the evolution of
  the low-$\beta$ region is reasonably understood by the above arguments.

\subsubsection{Boost of Explosion via MRI}
\begin{figure}[t]
\epsscale{1}
\plotone{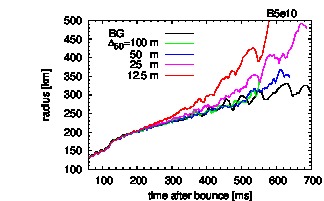}
\plotone{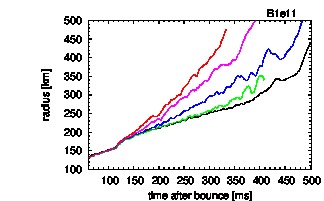}
\plotone{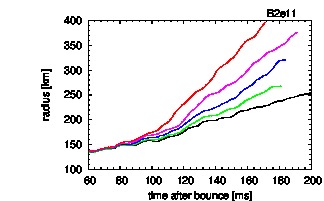}
\caption{Time evolutions of the shock radii at equator for all the models.}
\label{fig.t-req}
\end{figure}

\begin{figure}[t]
\epsscale{1}
\plotone{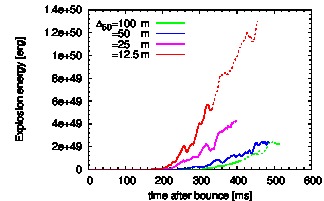}
\caption{Time evolutions of the diagnostic explosion energies for
  model series B1e11. After the maximum shock position exceeds the
  outer radial boundary, they are plotted by dotted lines.}
\label{fig.t-eexp}
\end{figure}

Although the variation of dynamical behavior with the
resolution seems rather complicated as described above, there is actually
one clear 
trend, i.e., the faster shock expansion at the equator for the
higher-resolutions (see 
Figure~\ref{fig.betamulti} and \ref{fig.t-req}). Since the equatorial region 
contains a larger amount of mass compared to the polar region, the larger
explosion energy is expected for the faster shock expansion. As shown
shortly, this is indeed the case. 

Figure~\ref{fig.t-eexp} shows the time evolution of the diagnostic
explosion energy, which is defined as the sum of the kinetic,
magnetic, internal, and gravitational energies over the fluid
elements that move outward with positive energies, for model series
B1e11. This clearly shows 
that the diagnostic explosion energy becomes larger as
the resolution is increased. 
Figure~\ref{fig.betamulti} indicates that the magnetic
effects are not necessarily lager for higher resolutions
(e.g., compare panel (e) and (h)),
which suggests that the magnetic pressure is not a key factor to boost
the explosion. Note that although the collimated jets are driven by
magnetic pressure, they give a minor contribution to the explosion energy 
due to their small volumes.
As pointed out by \cite{saw14}, the increase in the explosion
energy is attributes to the more efficient neutrino heating in 
higher resolution models.

\begin{figure}[t]
\epsscale{1}
\plotone{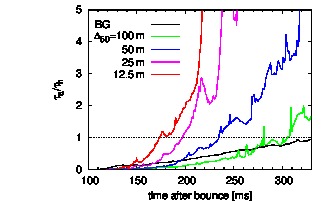}
\plotone{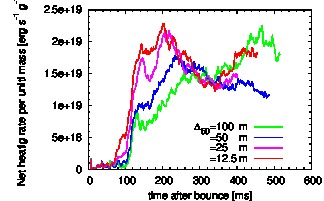}
\caption{Time evolutions of $\tau_{\textrm{a}}/\tau_{\textrm{h}}$
  (upper panel) and the net heating rate per unit mass averaged over
  the heating region (lower panel) for model series B1e11.
See text for the definition of $\tau_{\textrm{a}}$ and $\tau_{\textrm{h}}$.}
\label{fig.t-heat}
\end{figure}

\begin{figure*}[t]
\epsscale{1}
\plottwo{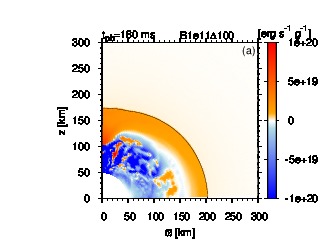}{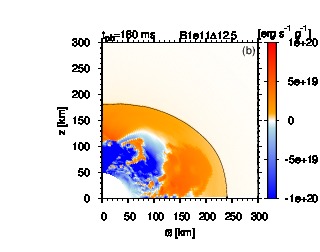}
\plottwo{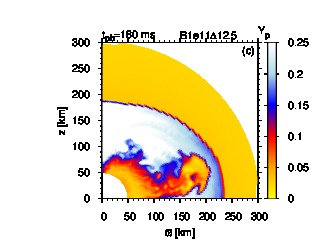}{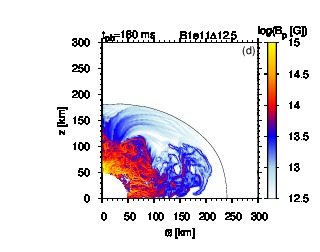}
\caption{Top panels: color maps for the net heating rate per unit mass
for models B1e11$\Delta$100 (panel (a)) and B1e11$\Delta$12.5 (panel
(b)) at $t_{\textrm{pb}}=180$~ms. Panel (c): Color map for the proton
fraction, $Y_p$, for model B1e11$\Delta$12.5. Panel (d): Color map for the
strength of the poloidal magnetic fields for model B1e11$\Delta$12.5.}
\label{fig.heat}
\end{figure*}

How close to revival the stalled shock is roughly measured by
the ratio of the advection timescale, $\tau_{\textrm{a}}$,
during which matter traverses the gain region, to the heating
timescale, $\tau_{\textrm{h}}$, within which matter gains enough
energy to overcome gravity \citep{tho00}. Following \citet{dol13}, we
define the advection timescale as
\begin{equation}
\tau_{\textrm{a}}=\int^{R_{\textrm{gain}}}_{R_{\textrm{sh}}}
\frac{dr}{\langle\langle v_r\rangle\rangle},
\end{equation}
where the double angle bracket implies that the solid-angle average
over $4\pi$ as well as the time average over the interval of 10~ms are taken. 
$R_{\textrm{sh}}$ is the mean shock radius, whereas
$R_{\textrm{gain}}$ is defined as the innermost radius at
which the solid-angle-averaged net heating is positive. The heating
timescale is defined as
\begin{equation}
\tau_{\textrm{h}}=
\frac{4\pi\int^{R_{\textrm{gain}}}_{R_{\textrm{sh}}}
\langle e+\frac{\rho v^2}{2}+\frac{B^2}{8\pi}+\rho\Phi\rangle r^2 dr}
{4\pi\int^{R_{\textrm{gain}}}_{R_{\textrm{sh}}}
\langle Q_E^{\textrm{em}}+ Q_E^{\textrm{abs}}\rangle r^2dr},
\end{equation}
where the single angle brackets mean that the only solid-angle average
is taken. 
The upper panel of Figure~\ref{fig.t-heat} plots the evolution of
$\tau_{\textrm{a}}/\tau_{\textrm{h}}$ for model series B1e11. The
comparison of this figure with Figure~\ref{fig.t-eexp}
indicates that shock revival, which is indicated by positive explosion
energies, roughly corresponds to
$\tau_{\textrm{a}}/\tau_{\textrm{h}}\gtrsim 1$. It is also evident that higher
resolutions result in higher heating efficiency. 

In \cite{saw14}, we
argued that this is due to the increase of $\tau_{\textrm{a}}$ in the
higher resolution models as a
result of more efficient angular momentum transfer, which leads to the
expansion of the heating region. This is true of the current models.
Comparison between models B1e11$\Delta$100 and
B1e11$\Delta$12.5 at $t_{\textrm{pb}}=180$~ms shows that the heating
region is thicker (see the upper panels of Figure~\ref{fig.heat}) and
the amount of angular momentum contained in the heating region is
larger for the latter model: they are
$7.0\times 10^{47}$g~cm$^{2}$s$^{-1}$ for model B1e11$\Delta$100
and $1.9\times 10^{48}$g~cm$^{2}$s$^{-1}$ for model
B1e11$\Delta$12.5 at $t_{\textrm{pb}}=180$~ms.

Besides the increment of $\tau_{\textrm{a}}$, we found in this paper
that the reduction of  $\tau_{\textrm{h}}$ owing to a larger heating
rate per unit mass is also contributing to the larger
$\tau_{\textrm{a}}/\tau_{\textrm{h}}$ in the higher resolution
models. As shown in the lower panel of Figure~\ref{fig.t-heat}, the
heating rate per unit mass during $\sim 100$--$250$~ms, the period
crucial to shock revival, becomes larger as
the resolution increases. The comparison of the upper panels of
Figure~\ref{fig.heat} for models B1e11$\Delta$100 and
B1e11$\Delta$12.5 indicates that this is originated in a patch of
region with large heating rates around the equator observed in the
latter model. We evaluated the heating 
and cooling rates separately and found that the relatively inefficient
cooling in the patch compared with the surroundings is responsible
for the larger net heating rate. From
Equation~(\ref{eq.qe.em}), the cooling rate per unit mass,
$Q_E^{\textrm{em}}/\rho$, is proportional to $(kT)^6$ and
$Y_n\mathcal{F}_5(-\eta_e)+Y_p\mathcal{F}_5(\eta_e)$. We found that
there is no
substantial difference in $(kT)^6$ between the patch and surroundings
but that $Y_n\mathcal{F}_5(-\eta_e)+Y_p\mathcal{F}_5(\eta_e)$ is several
times smaller in the patch. 
In the surroundings, where $Y_n\sim Y_p\sim 0.2$,
$\mathcal{F}_5(-\eta_e)\sim 10$, and $\mathcal{F}_5(\eta_e)\sim 900$,
the products are $Y_n\mathcal{F}_5(-\eta_e)\sim 2$ and
$Y_p\mathcal{F}_5(\eta_e)\sim 200$, viz., the cooling is dominated by
electron capture since electrons are much more abundant than positrons.
On the other hand, the electron capture is found to be relatively
inactive in the patch due to small number of protons ($Y_p\sim 0.05$, see
panel (c) of Figure~\ref{fig.heat}) and electrons
($\mathcal{F}_5(\eta_e)\sim 500$): the product is
$Y_p\mathcal{F}_5(\eta_e)\sim 30$. We found that the positron
capture rate is small as well in the patch:
$Y_n\mathcal{F}_5(-\eta_e) \sim 20$, where $Y_n\sim 0.8$ and
$\mathcal{F}_5(-\eta_e)\sim 30$. To summarize, the the larger heating
rate in the patch is caused by poverty of protons and electrons, and
a low electron capture rate as a consequence.

The low-$Y_p$ (equivalently low-$Y_e$) region coincides with the location
where the poloidal magnetic field is relatively strong (compare panels
(c) and (d)
of Figure~\ref{fig.heat}). This suggests that low-$Y_p$ fluids
originally located at small radii are drifted along the magnetic flux 
loops by MRI.  
We also found that the outflow along the rotation axis observed in model
B1e11$\Delta$100 and the collimated jets found in models
B1e11$\Delta$25 and B1e11$\Delta$12.5 (see panels (e), (k), and (n) of
Figure~\ref{fig.betamulti},
respectively) also convey low-$Y_p$ matter from deep inside the
core, which is reflected to the rise of the volume-averaged net heating
rate seen after 
$t_{\textrm{pb}}\sim 350$~ms for these models (see green, magenta, and
red lines in the bottom panel of Figure~\ref{fig.t-heat}).

\begin{figure}
\epsscale{1}
\plotone{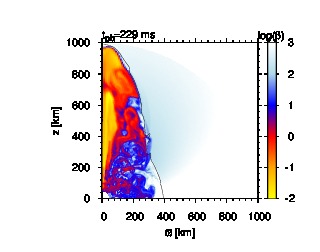}
\plotone{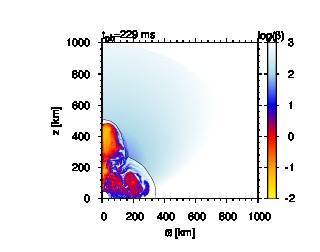}
\caption{Color maps for the plasma beta for models B2e11H (upper panel) and
  B2e11NH (lower panel) at $t_{\textrm{pb}}=229$~ms.}
\label{fig.beta1000}
\end{figure}

In order to estimate the possible influences of these effects on the
global dynamics, we carried out
two groups of additional test simulations based on models B1e11$\Delta$50 and
B2e11$\Delta$50. The first one of them are extends the radial outer
boundaries to $r=1000$~km (models B1e11H and
B2e11H). The other one is different in that the net neutrino heating
is switched off, i.e., 
$Q_E^{\textrm{abs}}+Q_E^{\textrm{em}}=\max\left[Q_E^{\textrm{abs}}+Q_E^{\textrm{em}},0\right]$
(models B1e11NH and B2e11NH). 

Figure~\ref{fig.beta1000} shows the profile of
the plasma beta for models B2e11H and B2e11NH at $t_{\textrm{pb}}=229$~ms, when 
the shock surface reaches $r=1000$~km in model B2e11H. Comparing
the two panels, we can immediately see the importance of the neutrino
heating: when the shock surface reaches $r=1000$~km in model B2e11H,
that in model B2e11NH has just passed the
radius of 500~km, and it arrives
at $r=1000$~km 65~ms later than model B2e11H. 
Even though the collimated jets observed in these 
two models are magnetically dominated, the neutrino heating plays a
significant role. The diagnostic explosion energies at 
the time when the shock fronts reach the radius of 1000~km are
$4.6\times 10^{49}$ and $1.8\times10^{49}$~erg, 
respectively, for models B2e11H and B2e11NH, implying that the
contribution of the neutrino heating to the explosion energy is
about 60\%. The neutrino heating is even more crucial for
weaker initial 
fields. The shock surface in model B1e11NH stays within
400~km even at 678~ms after bounce while that of
B1e11H has passed the radius of 1000~km at
$t_{\textrm{pb}}=595$~ms. The diagnostic 
explosion energy in model B1e11H is 
$4.7\times 10^{49}$~erg at the time when the shock front reach the
  radius of 1000~km, while that in model B1e11NH is negligibly small,
$\lesssim 2\times 10^{47}$~erg, through the simulation. Note that
  longer time simulations following the propagation of the shock front
  through the whole progenitor would be necessary to correctly measure
  the explosion energy.

The small contribution of magnetic field to the explosion
  energy discussed above is a substantial difference from previous MHD
  simulations involving magnetar-class initial fields, in which
  a magnetic field alone boosts the explosion energy up to
  $\sim10^{51}$~erg, accompanying a jet with a rather-large opening
  angle \citep[e.g.,][]{yam04}. This implies that in our models, the
  Maxwell stress is weaker, and thus the extraction of the rotational
  energy is less efficient than in those simulations.

\subsection{Effects of Inner Boundary}\label{sec.ibc}
We have seen that the MRI efficiently amplifies weak seed
magnetic fields to dynamically important strengths, having a
positive impacts on explosion. In this section, we investigate
whether the above results depends on the location of the inner
boundary, shifting it to smaller radii in model B2e11$\Delta$50.

We ran a new simulation with the inner boundary
at $r=30$~km to take the effect of strong differential
rotation beneath the radius of 50~km into account, which we refer
to as model Rin30. The  
spatial resolution of this model is similar to
that of model B2e11$\Delta$50 outside the 
radius of 50~km and is 30~m at the inner boundary.
The fraction of the volume where $N_{\textrm{MRI}}$ is less than 10 is only a few
percent inside the radius of 50~km, which is similar to that outside
(see Figure~\ref{fig.Nmri}).  

\begin{figure}[t]
\epsscale{1}
\plotone{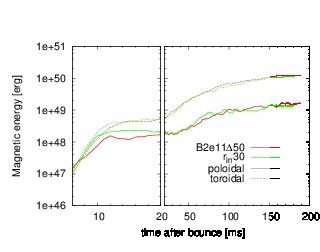}
\plotone{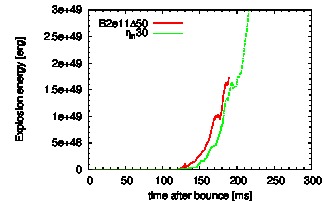}
\caption{Upper panel: time evolutions of the magnetic
energies contained in the range of $50<(r/$km$)<500$ for models
Rin30 and B2e11$\Delta$50. Lower panel: time evolutions of the
diagnostic explosion energies for models Rin30 and B2e11$\Delta$50. After
the maximum shock position exceeds the outer radial boundary, they
are plotted by dotted lines.}
\label{fig.ibc}
\end{figure}

The growth rates of MRI inside the radius of 50~km are found to be
lager than those outside on average. The upper panel of
Figure~\ref{fig.ibc} indicates that the exponential growth rate
of $E_{B_p}$ averaged over the range of $50<(r/$km$)<500$ is larger
for model Rin30 than for model B2e11$\Delta$50. This implies that the
magnetic fields amplified inside the radius of 50~km are advected
outward in model Rin30. Note that the degree of differential
  rotation is even greater inside the radius of 30~km, and thus the
  above effect would be more pronounced if we carried out the
  simulation with a yet smaller radius of inner boundary.
On the other hand, we found that the saturation level of
$E_{B_p}$ is unaffected by the position of the inner boundary, which
may be reasonable if the saturation is determined
by the strength of the numerical diffusion, and hence the spatial
resolution, which are more or less the same (see
Section~\ref{sec.mri}). The evolution of $E_{B_t}$ is similar to that
of $E_{B_p}$. 

The global dynamics does not change significantly, either, by moving the
inner boundary position. The low-$\beta$ collimated jet that emerges from
inside the roughly-spherical shock is a feature common to both models
B2e11$\Delta$50 and Rin30. Whereas the evolutions of the jets are rather
different between the two models, the shock propagation on the equator
are very similar between the two. As shown 
in the lower panel of Figure~\ref{fig.ibc}, the diagnostic explosion
energies of the two models are also nearly identical. 
From these results, we believe that the conclusions of the
current study are not affected by our choice of the inner boundary
position, $r=50$~km.

\section{Discussion and Conclusion}\label{sec.discon}
We have performed MHD simulations of the core-collapse of rapidly-rotating
magnetized stars in two dimensions under axisymmetry, changing both the
strength of magnetic field and the spatial resolution.
Our goal is to study the behavior of the MRI in core-collapse
supernovae and its impacts upon the global dynamics.
As a result of computations we found the followings.

The MRI greatly amplifies the seed magnetic fields even in the
dynamical background of the core-collapse. Although the dominant mode,
buoyant mode or Alfv\'en 
mode, differs from location to location, the former plays a primary
role in the exponential growth phase. It is true that the linear
growth rate of the buoyant mode is independent of the wavelength, a
certain degree of high spatial resolution seems necessary to correctly
capture the exponential 
growth. The magnetic energies of the poloidal component gets nearly
saturated within the simulation times in all models,
where the saturation level is higher for larger initial magnetic
fields. The magnetic energies of the toroidal component grow
continuously in most of the models, on the other hand, and the core
becomes toroidal-field dominant. 

The MRI also has a grate impacts on the global dynamics.
Models in which the MRI are well resolved show faster expansions of
shock surface and obtain more powerful explosions. The formation of 
collimated jet is also found in models where the initial magnetic
field is relatively strong and the MRI is well resolved.
The following two effects are found to be the key to the boost of
explosion: the first one is the expansion of the heating region due to
the outward angular momentum transfer. This makes the advection
timescale, or the time for matter to traverse the heating region, longer and
thus enhance the heating; the second effect is the drift of
low-$Y_p$ (equivalently low-$Y_e$) matter along the MRI-distorted
magnetic flux loops as well 
as their ejection by the jets, from deep inside the core to the heating
region. The cooling due to electron capture is reduced in the low-$Y_p$
region, and the net heating rises as a result.

The diagnostic explosion energies obtained in the current simulations
are much smaller than the typical value of $\sim 10^{51}$~erg in
reality. Note, however, that our choice of the neutrino
luminosity, $L_{\nu_e}=L_{\bar{\nu_e}}=1.0\times10^{52}$~erg~s$^{-1}$,
which is assumed to be constant,
is quite modest. According to the core-collapse simulations by 
\citet{bru13}, who employed the flux-limited diffusion with the
ray-by-ray-plus approximation for neutrino transport, both
$L_{\nu_e}$ and $L_{\bar{\nu_e}}$ are 
$\sim 5.0\times 10^{52}$~erg~s$^{-1}$ at $t_{\textrm{pb}}\sim 100$~ms
and decay to $\sim 1.0\times 10^{52}$~erg~s$^{-1}$ over several hundred
milliseconds. If such an evolution of luminosity is adopted in our
simulations more energetic explosion would be obtained.

Although our choice of the initial magnetic field strength, $\sim
10^{11}$~G, is much smaller than those assumed in the former global
MRI simulations 
\citep[e.g.,][]{obe06,shi06,saw13a}, most of progenitors of
core-collapse 
supernovae may posses even weaker magnetic fields \citep{wad14}. 
Since lower saturation magnetic fields are expected for
weaker initial magnetic fields according to our results, the impact of
MRI in "normal'' 
supernovae will be smaller than that found in this work. Although it
is important to study much weaker
magnetic fields, simulations will be computationally expensive and
thus currently unfeasible: a reduction of the initial magnetic field by half,
with the spatial resolution kept at the current level, demands
eight times higher computational cost for 2D time-explicit simulations.

The dependence on the initial rotation also needs to be investigated,
since the rotation speed of stars is also likely to 
distribute over a wide range \citep{ram13}. We are currently undertaking
such studies, and the results will be presented elsewhere in the future.

Although our simulations are 2D under axisymmetry, supernovae
occur in three dimensions in reality, and non-axisymmetric effects such
as dynamo, three-dimensional turbulence, non-axisymmetric modes
  of various instabilities may be important. One should
  keep in mind that these effects possibly alter results obtained
  by current 2D simulations. For example,
3D-MHD simulations performed 
by \citet{mos14} demonstrated that magnetically driven-jets can be
destroyed by the $m=1$ mode of a kink-type instability, whereas
such destruction of the jet was not observed in 3D-MHD simulations by
\citet{mik08}. In order to know how essential non-axisymmetric
  effects are, 3D global simulations are 
mandatory. During the reviewing process of this paper,
  \citet{mos15} published the results of the first global-3D
simulations of the MRI in proto-neutron stars. Under
quadrant symmetry, they had simulated the evolution of the MRI
for 10~ms, and found the formation of large-scale, strong toroidal
fields, which hints at later magnetically-driven mass ejections. Such
simulations have only just begun, and the possible 3D effects mentioned above
should be studied in detail in the future.
This requires long-term, large-domain, full 3D
simulations, which may be
marginally feasible with exa-flops computers of the next generation.

\citet{mas07} and \citet{gui15} argued that the neutrino viscosity may
hamper the 
growth of MRI deep inside the core, i.e.,
$r\lesssim 30$~km for a fast rotation like ours. Applying the
magnetic field of $\sim 10^{13}$--$10^{14}$~G 
obtained in our BG runs for $r\lesssim 30$~km to the 
fast rotation model of \citet{gui15}, we found that the neutrino
viscosity may be 
marginally important there. Since the inner boundary
condition of our MRI runs is 
given by the data of the BG runs of low resolution, the artificial
suppression of MRI by numerical diffusions may effectively mimic the
damping by the neutrino viscosity. We hence believe that
full-sphere simulations including the neutrino viscosity
will not change our conclusions in this paper so much, if the viscous
process is important at all.

\acknowledgments
H.S. is grateful to Kenta Kiuchi, Nobuya Nishimura, Yuichiro
Sekiguchi, and Tomoya Takiwaki for fruitful discussion.
H.S. also thank Daisuke Yamaki and Hideki Yamamoto at RIST, Kobe
Center for useful advice about MPI and openMP parallelization.
Numerical computations in this work were carried out on Cray XC30 at
Center for Computational Astrophysics, National Astronomical
Observatory of Japan, and on HITACHI SR16000 at the Yukawa Institute
Computer Facility.
This work is supported by a Grant-in-Aid for Scientific Research from
the Ministry of Education, Culture, Sports, Science and Technology,
Japan (24103006, 24244036, 26800149).

\end{document}